\documentclass[%
 twocolumn,
superscriptaddress,
nofootinbib,
 amsmath,amssymb,
 aps,
longbibliography,
]{revtex4-2}

\usepackage{etoolbox}
\usepackage{graphicx}% Include figure files
\usepackage{dcolumn}% Align table columns on decimal point
\usepackage{amsmath}
\usepackage{bm}% bold math
\usepackage[%
  colorlinks=true,
  urlcolor=blue,
  linkcolor=blue,
  citecolor=blue
]{hyperref}% add hypertext capabilities

\usepackage{amsmath}
\usepackage{amssymb}
\usepackage{bm}

\newcommand{\taur}{\tau_{\rm r}}

\newcommand{\Pe}{{\rm Pe}}
\newcommand{\epsw}{{\varepsilon_{\rm w}}}
\newcommand{\epswlj}{{\varepsilon_{\rm w}^{\rm LJ}}}
\newcommand{\Iell}{I_{\ell}}

\begin{document}

%%%TITLE, AUTHORS AND ABSTRACT%%%
\title{Partial and complete wetting of droplets of active Brownian particles}

\author{Francesco Turci}
\affiliation{H. H. Wills Physics Laboratory, University of Bristol, Tyndall Avenue, Bristol BS8 1TL, United Kingdom}

\author{Robert L. Jack}
\affiliation{Department of Applied Mathematics and Theoretical Physics, Centre for Mathematical Sciences, University of Cambridge, Wilberforce Road, Cambridge CB3 0WA, United Kingdom}
\affiliation{Yusuf Hamied Department of Chemistry, University of Cambridge,
Lensfield Road, Cambridge CB2 1EW, United Kingdom}

\author{Nigel B. Wilding} 
\affiliation{H. H. Wills Physics Laboratory, University of Bristol, Tyndall Avenue, Bristol BS8 1TL, United Kingdom}

\begin{abstract}
We study wetting droplets formed of active Brownian particles in contact with a repulsive potential barrier, in a wedge geometry.  Our numerical results demonstrate a transition between partially wet and completely wet states, as a function of the barrier height, analogous to the corresponding surface phase transition in passive fluids.
 We analyse partially wet configurations characterised by a nonzero contact angle $\theta$ between the droplet surface and the barrier, including the average density profile and its fluctuations. These findings are compared with two equilibrium systems: a Lennard-Jones fluid and a simple contour model for a liquid-vapour interface.  We locate the wetting transition where $\cos(\theta)=1$, and the neutral state where $\cos(\theta)=0$.  
We discuss the implications of these results for possible definitions of surface tensions in active fluids.
\end{abstract}

\maketitle

%%%MAIN TEXT%%%%

%% latex macros for notation

\newcommand{\gama}{ \tilde\gamma_{\rm lv} }
\newcommand{\gamb}{ \Delta\tilde\gamma_{\rm w} }

\section{Introduction}

Active matter systems display a wide range of surprising phenomena in their non-equilibrium steady states.  Among the simplest active systems are fluids comprised of self-propelled particles without aligning interactions, which are known as scalar active matter~\cite{Wittkowski2014,solon2018,speck2022}.
 Many such systems exhibit motility-induced phase separation (MIPS)~\cite{cates2015motility}, which resembles equilibrium coexistence of dense and dilute fluid phases.  However, the interfacial properties of these active fluids phases differ significantly from their equilibrium counterparts which can lead -- for example -- to microphase separation and bubbly phases~\cite{Tjhung2018}.

Given such observations, it is natural to ask about other interfacial properties of active fluids, and their similarities and differences with equilibrium systems.  An interesting example occurs when a system undergoing MIPS is placed in contact with a solid or penetrable substrate (or ``wall'').  In this case, one may expect analogues of the rich phenomenology of wetting, as it occurs in equilibrium fluids at liquid-vapor coexistence~\cite{degennes1985}.  Processes reminiscent of equilibrium wetting appear to play a crucial role in active systems composed of living cells, soft responsive materials, and embedded energy sources. For example, wetting and dewetting on soft substrates enables tunable adhesion, motility, and shape change of cells \cite{brugues2014,alert2020,bhattacharjee2022,pallares2023} and control the motility of bacteria at interfaces \cite{harshey2003}.

In equilibrium, wetting behaviour can be analysed in several different settings.  A famous example is the formation of a liquid droplet on a weakly attractive solid substrate.  The contact angle $\theta$ of this droplet obeys Young's equation: 
\begin{equation}
   \gamma_{\rm lv} \cos\theta= \gamma_{\rm wv}-\gamma_{\rm wl} \:,
    \label{eq:young}
\end{equation}
where $\gamma_{lv} $ is the liquid-vapor surface tension and similarly $ \gamma_{\rm wv}, \gamma_{\rm wl}$ are surface tensions between the fluid phases and the wall.  Increasing the attraction between the fluid and the wall, the tension $\gamma_{\rm wl}$ decreases and the droplet spreads out, leading eventually to a wetting transition~\cite{bonn2009,bonn2001}  as $\theta\to0$ (specifically, this is the transition from partial to complete wetting, but we term it here ``the wetting transition'', for simplicity).  This transition -- and related phenomena such as drying transitions~\cite{evans2015,evans2017} -- may be either first-order or critical, depending on the behaviour of $\cos\theta$ as the transition is approached.
Measurements of contact angle thus permit the characterisation of wetting transitions on planar surfaces \cite{ingebrigtsen2007} as well as related situations such as the filling transitions that occur in capillaries \cite{malijevsky2021}.

For computational model fluids like Lennard-Jonesium, measurement of contact angles tends to be challenging.  However, there are convenient alternative approaches which either exploit the grand canonical ensemble, or a slit geometry with a fluid confined between two walls (and periodic boundaries in the other direction).  Droplets do not form in these cases, and one instead focusses on the average density profile $\rho(z)$, as a function of the distance $z$ from the wall.  For a grand canonical system with a single wall, one defines the adsorption $\Gamma \equiv \int_0^{\infty} d z\left(\rho(z)-\rho_b\right)$ where $\rho_b$ is the bulk density at $z\to\infty$.  This quantity is accessible experimentally~\cite{law2001,gatica2009} as well as in density functional theories~\cite{evans2017,evans2019} and in simulations~\cite{evans2019,nijmeijer1990} (with the aid of finite-size scaling). Increasing the attraction between the fluid and the wall, a wetting transition occurs when $\Gamma$ becomes infinite, which may occur either by a smooth divergence (critical wetting) or by a discontinuous jump (first-order wetting), depending on the range of fluid-fluid and wall-fluid interactions.  A similar analysis can be performed for the slit geometry in the canonical ensemble, in which case the wetting transition is signalled by a symmetry breaking of the density profile $\rho(z)$ with respect to the two walls \cite{nijmeijer1990,turci2021a}.  

In all these equilibrium cases, statistical mechanical theories place strong constraints on the phenomenology.  For example, the surface tensions in \eqref{eq:young} can be defined unambiguously through gradients of an appropriate free energy, as can the adsorption $\Gamma$~\cite{evans2017,evans2019}.  This provides consistency requirements between different ensembles and geometrical settings: 
studies based on the adsorption and the contact angle both deliver the same results for the locations and properties of surface phase transitions, as long as finite-size effects are controlled.

By contrast, active fluids are not ruled by a free energy, and the status of their wetting transitions is much less well-understood.  
Indeed, there are several different proposals for active generalisations of the liquid-vapour surface tension~\cite{speck2020a,omar2020,lauersdorf2021,hermann2021a,chacon2022,li2023,bialke2015,zakine2020,fausti2021,langford2023,omar2023}.  It is not clear \emph{a priori} whether any suitably generalised version of Young's equation should apply for these systems; if some such generalisation does exist then one may ask
which (if any) of the liquid-vapour surface tensions might appear, and what should be used in place of $\gamma_{\rm wv}-\gamma_{\rm wl}$.

In recent work by some of us~\cite{turci2021a}, an approach based on the absorption $\Gamma$ was used to analyse the wetting properties of a paradigmatic active fluid, comprised of active Brownian particles (ABPs) in $d=2$ and $d=3$ dimensions.  An important difference from equilibrium fluids is that active particles tend to accumulate at walls, even in the absence of attractive interactions~\cite{speck2016a,elgeti2013}.   As a result, an infinitely repulsive `hard' wall is always wet for these active fluids, in contrast to equilibrium fluids for which  a hard wall remains dry ~\cite{Henderson:1985aa}.  However, on replacing a solid wall with a penetrable barrier, behaviour similar to first-order wetting was found in $d=3$, for the slit geometry~\cite{turci2021a}.

In this work, we take a complementary approach to Ref.~\citenum{turci2021a}, which is to examine the wetting behaviour of droplets in the same $3d$ system of ABPs.  Such studies are numerically challenging due to finite-size effects which appear in the form of large fluctuations of the droplet shape and position.  We show that this can be mitigated by confining droplets in a wedge geometry.  
We identify a discontinuous wetting transition for droplets (where $\cos\theta\to1$).  The nature and location of this transition are consistent with the slit geometry.  We also characterise the situation of neutral wetting ($\cos\theta=0$).  We give a critical analysis of Young's equation in this setting.  We argue that by defining surface tensions in terms of the probabilities of droplet shape fluctuations, Young's equation holds \emph{by definition} for the most likely shape, as long as shape fluctuations are controlled by local properties of the interface.  In this case, the liquid-vapour surface tension also determines the probability of large-wavelength capillary waves.

Our paper is organised as follows: the ABP model is defined in Sec.~\ref{sec:abp-model} and its wetting behaviour is discussed extensively in Sec.~\ref{sec:abp-wet}.  Then Sec.~\ref{sec:lj} discusses analogous behaviour in a passive LJ fluid.  Sec.~\ref{sec:contour} introduces the simple contour model for equilibrium interfacial fluctuations and compares it with the particle-based models.  These results are discussed in Sec.~\ref{sec:conc}, which also summarises our main conclusions.

\section{Model and geometrical setup}
\label{sec:abp-model} 

We analyse ABPs as a prototypical model of scalar active matter
that displays MIPS in the bulk~\cite{fily2012athermam,redner2013}. 
We follow a previous parameterization \cite{stenhammar2014} described in Appendix~\ref{sec:model-details}, with ABPs interacting via the Weeks-Chandler-Anderson potential of lengthscale $\sigma$, and energy scale $\varepsilon$, with coupled translation and rotational diffusion constants $D_r=3D_t/\sigma^2$ and self-propulsion velocity $v_0$ defining a P\'eclet number $\Pe=v_0/\sigma D_r$.

We simulate this model in a fully periodic orthorhombic box of dimensions $L_x\times L_y\times L_z$. We consider various box sizes, but we typically elect to work with $L_z=20\sigma$, smaller than both $L_x$ and $L_y$. 
In order for a surface phase transition to occur, the system must be at a state point for which bulk liquid-vapor coexistence occurs.  Accordingly, we choose model parameters well inside the MIPS region, specifically number density $\rho=0.60$ and constant P\'eclet number $\Pe = 60$ (see \cite{turci2021} for the bulk phase diagram).   
Applying the lever rule, this corresponds to an approximate liquid fraction of the system $f=(\rho-\rho_{\rm LD})/(\rho_{\rm HD}-\rho_{\rm LD})\approx(0.6-0.45)/(1.25-0.45)=0.1875$, 
which results in a cylindrical liquid domain with its axis parallel to the $z$ axis. Such a geometry allows us to monitor the changes in the liquid-vapour interfaces (and, in particular, the contact angle) via its two-dimensional projections in the $x-y$ plane.  We note that while MIPS is unstable with respect to vapor-crystal phase separation in this model \cite{turci2021,omar2021}, crystallisation is not observed on our simulation timescales.

\begin{figure*}[t!]
    \centering
    \includegraphics[width=\textwidth]{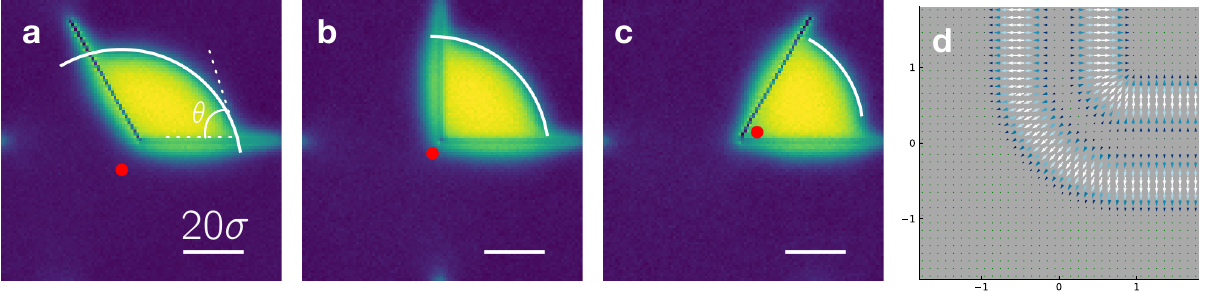}
    \caption{
    (a-c) Droplet configurations for three wedges of aperture angles $\alpha=120^{\circ},90^\circ,60^\circ$ respectively, all at barrier strength $\epsw=12$. Each geometry promotes occupation of the wedge by droplets of active Brownian particles that share similar contact angles ($\theta=75^\circ\pm3^\circ $, $83^\circ\pm4^\circ$, $85^\circ\pm 9^\circ$ respectively). The fitted circular arcs (white continuous lines), with their centres (red dots), are also plotted. (d) The shape of the barriers' static force field around the corner of the wedge. 
    }
    \label{fig:apertures}
\end{figure*}

To induce wetting, we introduce a penetrable wall, which we model as a static repulsive barrier\cite{turci2021a}.
To this end, we identify a piecewise surface {$\mathsf{S}$}
and employ an external repulsive potential perpendicular to the surface, which takes the form of a cosine hump: 
\begin{align}
V(r) = 
\begin{cases} 
\epsw[\cos (\pi r / d)+1] 
    & \text{if } r < d \\
&\\  % blank row
0
    & \text{otherwise }
\end{cases}
\label{eq:barrierpot}
\end{align}
where $r$ is the perpendicular distance from the surface $\mathsf{S}$. We take $d=\sigma$, which corresponds to a short-ranged (ie. thin) potential barrier. 

Simulations of liquid droplets of ABPs are challenging because of large fluctuations of the droplet shape.  This can be mitigated by increasing the system size, but the dilute (vapor) phase of MIPS has a relatively high concentration $\rho_{\rm LD}\approx 0.45$, which means that such simulations quickly become expensive, involving very large numbers (hundreds of thousands) of particles.  While planar walls are natural for wetting, we have found it helpful to choose $\mathsf{S}$ to take a wedge shape; this localises the droplet (reducing fluctuations) and accelerates its nucleation.  
 To achieve this, we choose $\mathsf{S}$ to comprise two finite planes that are joined together at an angle $\alpha$ by a cylindrical section, which gives the wedge a rounded corner. Fig.~\ref{fig:apertures}(a-c) illustrate the resulting setup, showing the planar density $\rho(x,y)$ (details of its numerical estimation are given below).  The $x-y$ force field generated by the barrier is illustrated in Fig.~\ref{fig:apertures}(d). Typical simulations have $L_x=L_y=100\sigma$ and $L_z=20\sigma$ with the wedge occupying one quadrant of the box. At the density considered, this results in simulations of $N=120\ 000$ particles. Working in three dimensions ensures that the density fluctuations are more controlled than the two-dimensional case and connects to previous evidence for a wetting transition that becomes sharper in the large $N$ limit \cite{turci2021a}.

\section{Wetting Phenomenology for Active Brownian Particles}
\label{sec:abp-wet}

\subsection{Average local density profiles: contact angles and wetting transition}

\begin{figure*}[t]
    \centering
    \includegraphics[width=\textwidth]{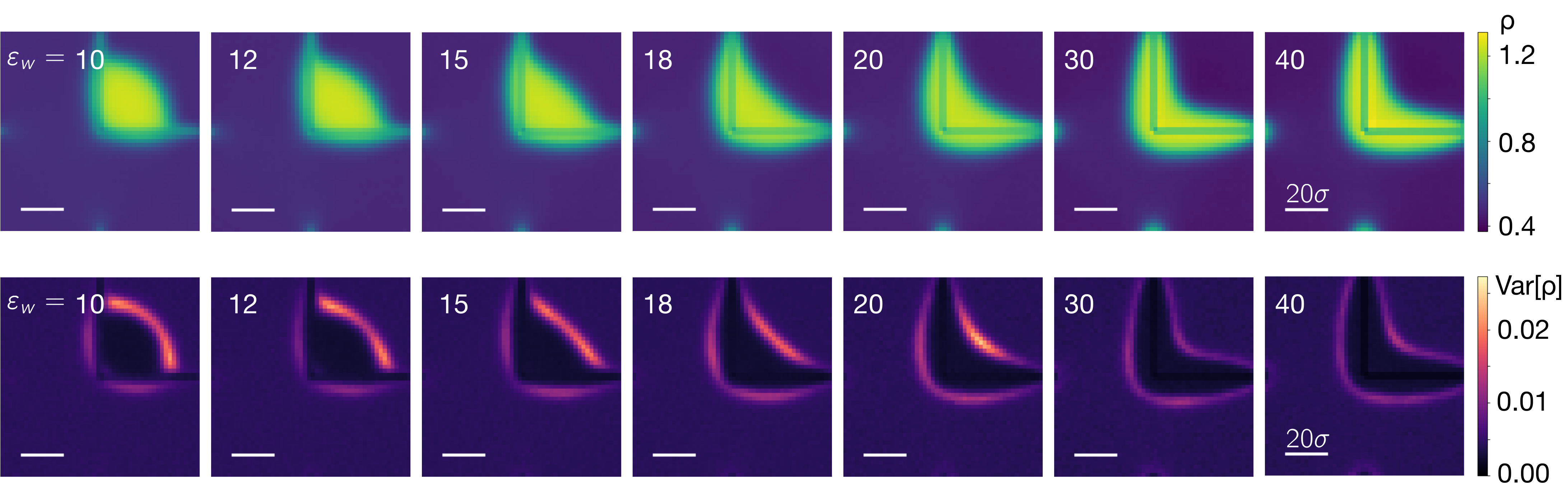}
    \caption{Average and variance of the local density in the active system. Upper row: stationary density profiles for phase-separated active Brownian particles in contact with a finite wedge-shaped potential of repulsive strength $\epsw$ for increasing values of the barrier strength in the range $\epsw=10-40$. The system has periodic boundary conditions. Lower row: stationary profiles of local density fluctuation ${\rm Var}[\rho]$ for phase-separated active Brownian particles at increasing values of the barrier strength $\epsw$.}
    \label{fig:densities-var}
\end{figure*}

Our analysis of wetting is based on the local density profile, which is obtained by
discretizing space in subvolumes of side $\ell=2\sigma$. We define the instantaneous number of particles in bin $j$ of center $\bm{b}_j$
\begin{equation}
    N_j = \sum_{i=1}^N \Iell(\bm{r}_i-\bm{b}_j),
\end{equation}
where $\Iell(\bm{x})$ is 1 if $x$ is within the cube of side $\ell$ centered at the origin and zero otherwise. The instantaneous number density at site $j$ is defined as
\begin{equation}
 \rho_j =  N_j/\ell^3,
 \label{equ:rhoj}
\end{equation}
and we average along the $z$ dimension to obtain two-dimensional density maps $\rho(x,y)=L_z^{-1}\int_0^{L_z}\rho(x,y,z)dz$. Following a relaxation time of  $100\taur$,  the average local density $\overline{\rho}(x,y)$ is formed by sampling at intervals of $\taur$, and averaging over a period of 1500$\taur$. 

Fig.~\ref{fig:apertures} shows two-dimensional projections of the density map for three different wedge aperture angles $\alpha$. The density profiles are smooth and -- for the value of $\epsw=12$ studied-- exhibit a liquid drop confined within the wedge. From such profiles one can, in principle, extract the vapor-liquid interface and estimate the contact angle $\theta$ between the active droplet and the repulsive barrier. For our system, the macroscopic notion of sharp contact between a circular vapor-liquid interface and the barrier is somewhat blurred by the ubiquitous presence of a thin layer of particles all around the wedge. To deal with this, we work with estimates of an \textit{apparent} contact angle, defined by fitting a circular arc to circular regions of the density profile as described further in Appendix~\ref{sec:contactangle}.  If a non-equilibrium analogue of Young's equation applies to these systems (valid in the large-system limit), then $\theta$ should be determined only by local properties of the three-phase contact line, so that $\theta$ is independent of $\alpha$. For the three value of $\alpha$ shown in Fig.~\ref{fig:apertures}, we find $\theta\approx 75^\circ-85^\circ$: this relatively small range seems consistent with the applicability of Young's equation. Note that for small equilibrium droplets Young's equation is modified by line tension and by Tolman corrections to the Laplace pressure \cite{toshev1988,widom1995,dobbs1999}, while in the active case curvature effects on swim pressure of confined ABPs have also been reported~\cite{smallenburg2015}; we explicitly neglect these effects in this work.

Henceforth we elect to work at fixed aperture $\alpha=\pi/2$. Fig.~\ref{fig:densities-var} illustrates how the stationary average density field $\overline{\rho}(x,y)$ for this value of $\alpha$ varies with the barrier strength $\epsw$.  At the largest values of $\epsw$ studied, thick liquid layers are present on both the interior and exterior of the wedge and the density profile flattens close to the corner of the wedge and terminates with a rounded shape at the tips. On lowering the barrier height from $\epsw=18$ to $\epsw=15$, liquid progressively accumulates in the wedge interior.  At $\epsw\approx 18$ the curvature of the interface between the interior liquid and the vapor changes sign from positive to negative and a recognisable droplet forms within the wedge having an apparent contact angle $\theta<\pi/2$. The fitting procedure to obtain $\theta$ is appropriate only when the density profile exhibits such a region of negative curvature,  ie. for $\epsw\lesssim 18$.  We assert (and confirm via a comparison with equilibrium wetting in a comparable geometry- see Secs.~\ref{sec:lj} and \ref{sec:contourcompare}) that the change in sign of the interfacial curvature corresponds to the transition from partial to complete wetting. In other words that the barrier is partially wet, (with $\theta>0$) for $\epsw\lesssim 18$, and is completely wet (ie. $\theta=0$) for $\epsw\gtrsim 18$. Within the partially wet regime, most liquid resides in the drop in the wedge interior. However some liquid resides on the exterior wall of the wedge forming a pair of symmetrical ``lobes'' of liquid-like density. A $\epsw$ decreases,  progressively more liquid accumulates in the drop whose contact angle $\theta$ increases, while the extent of the lobes decreases. In Sec.~\ref{sec:fs} we discuss the finite-size scaling behaviour of the lobes and the liquid drop. 

At $\epsw=10$ we find $\theta\approx\pi/2$, which - in the context of Young's equation - is interpreted as a \textit{neutral point}, where the tensions between the barrier and liquid and the barrier and the vapor balance each other. This point separate the partial wetting and partial drying regimes. On decreasing the barrier strength still further, we find that within the timescale of our simulations, the stationary state becomes harder to define: if we initiate the system in a homogeneous density state, the nucleation of the droplet becomes very slow; if we start instead from a pre-formed droplet and instantaneously decrease the barrier strength, the droplet progressively detaches from the barrier. The detachment indicates that the vapor phase is favoured near the barrier in the stationary state. Henceforth, we restrict our analysis to the range of barrier strength for which the liquid is attached to the barrier, where surface physics plays a role.

\begin{figure}
    \centering
    \includegraphics[scale=1]{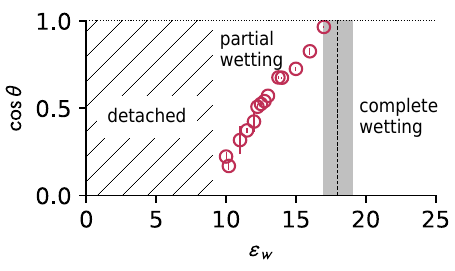}
    \caption{Cosine of the apparent contact angle $\theta$ for ABPs in a $90^\circ$  wedge for a range of repulsive strengths $\epsw$, as calculated from a circular arc fit to the portion of the liquid-vapor interface having negative curvature. The vertical dashed line indicates the transition from partial to complete wetting determined previously via a different approach \cite{turci2021a}. The hatched area indicates the regime in which the droplet detaches from the barrier.
    Error bars are bootstrap estimates for three standard deviations.
    }
    \label{fig:active-cosines}
\end{figure}

Our results for the dependence of the measured contact angle on the barrier strength are displayed in Fig.~\ref{fig:active-cosines}. For the weakest accessible barrier strengths at which the droplet is attached to the wedge, we find $\cos(\theta)\approx 0$, i.e. the droplet is close to the neutral point. As $\epsw$ increases, $\cos(\theta)$ increases approximately linearly and appears to attain the wetting point $\cos\theta=1$ with a nonzero slope, suggesting a first-order wetting transition around $\epsw= 18\pm 1 $.  This value of the wetting point accords with our previous analysis of this system in the slit geometry~\cite{turci2021a}.

\subsection{Local density fluctuations}

Density fluctuations play a central role in characterising phase transitions, both in passive and active systems. In the case of surface phase transitions studied within a grand canonical framework, the behaviour of the local compressibility (which quantifies the magnitude of the density fluctuations relative to the average density) allows the order of a surface phase transition \cite{evans2017} to be ascertained and provides information on the character of interfaces.  

To quantify the density fluctuations in our system, we accumulate the local variance of the density field ${\rm Var}[\rho (x,y) ]= \langle\rho^2(x,y)\rangle-\langle \rho(x,y)\rangle^2$ on the same scale $\ell=2\sigma$ over which the field is defined. This provides a fine-grained description of the spatial dependence of the density fluctuations. Fig.~\ref{fig:densities-var} shows that the fluctuations are greatest at the liquid-vapor interface: both for the droplet in the wedge interior and for the exterior lobes. In particular, the complete wetting regime at large $\epsw$ corresponds to small overall fluctuations, with higher values occurring around points of higher curvature.  As the repulsive barrier gets weaker (moving from right to left in Fig.~\ref{fig:densities-var}), the fluctuations at the liquid-vapor interface in the interior increase in magnitude, and become progressively more localised in the vicinity of the contact region between the interface and the barrier.  

\begin{figure*}[t]
    \centering
    \includegraphics{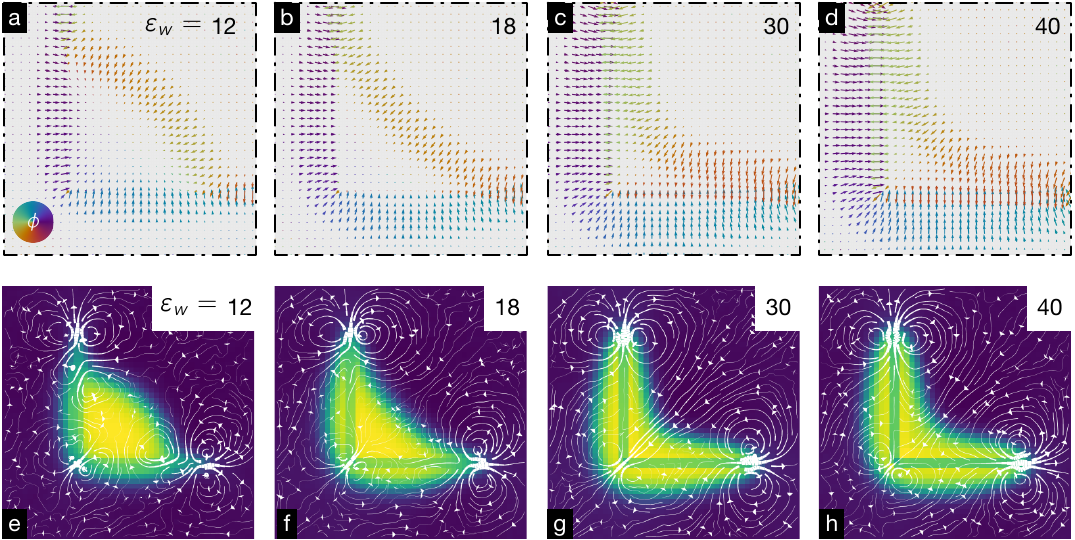}
    %[width=\columnwidth]{data/polarisation/composition-polar.pdf}
    \caption{Polarisation and flow. (a-d) Stationary average of the local particle orientation $\phi$ for increasing values of the barrier strength, accumulated on a grid of bin size $\ell=2\sigma$. The color coding indicates the angle of the local orientation; the arrow length corresponds to the average value of the polarization at the grid point, and it is scaled for clarity (leading to some crossings where the polarization is largest). For clarity, the plots depict only the immediate vicinity of the droplet, with grid binning $\delta x=2\sigma$. Both the dilute phase and the interior of the droplet have negligible average polarization: only the interfaces (between the liquid and the vapor and between the barrier and the fluid phases) display significant contributions orthogonal to the interfaces. (e-h) Streamlines of the particle flux obtained as described in the text for the indicated values of the repulsive barrier strength $\epsw$ overlayed to the density field. The line thickness scales linearly with the norm of the two-dimensional flux, with the same scaling used for all plots. The streamplots visualise the entire x-y span of the simulations.}
    \label{fig:polarisation-flux}
\end{figure*}

\subsection{Polarisation field}

The orientation of the self-propulsion force of ABPs rotates via a diffusive process. For a bulk system, every particle orientation is equally likely, and thus, the average orientation vector of each particle is zero.  When interfaces are formed (as in MIPS), the local stationary orientation can take nonzero values, indicating a local polarisation of the system. 

We define the instantaneous polarisation $\bm{\phi}$ from the particle orientation $\bm{n}_i$ on a grid of spacing $\ell$. For bin $j$ 

\begin{equation}
    \bm{\phi_j} =  \frac{1}{N_j}\sum_i \bm{n}_i \Iell(\bm{r}_i-\bm{b}_j),
\end{equation}
 which corresponds to the local average of the orientation field. Its norm is $|\bm{\phi}|\in[0,1]$, and it is expected to be negligible in the bulk phases, as the individual orientations diffuse on the unit sphere. Near boundaries and interfaces, however, the net orientation typically follows the density gradient, i.e. normal to the interface.

In Fig.~\ref{fig:polarisation-flux}(a-d), we track the changes to the stationary polarisation that occur as we vary the barrier strength $\epsw$.  As expected, the polarisation fields are non-zero only at interfaces and closely follow the density gradient, directed from the dilute towards the denser phase. The principal effect of varying $\epsw$ is manifest in the barrier region: For large $\epsw$, the particles at liquid-like densities on both the interior and exterior of the wedge are oriented against the barrier. For weaker barriers (e.g. $\epsw=12$), the particles inside the droplet change orientation and point towards the droplet interior. This contrasts with the behaviour of exterior particles in the vicinity of the barrier, which are always polarised towards the barrier, independent of $\epsw$.

\subsection{Finite size effects}
\label{sec:fs}

The density profiles display the formation of both an interior liquid layer and some accumulation on the exterior of the wedge. These exterior density ``lobes'' are a consequence of the non-equilibrium accumulation mechanism and its coupling with the barrier strength. In the limit of very repulsive barriers, the liquid covers both the interior and exterior approximately symmetrically, whereas, in the partial wetting regime that occurs for $\varepsilon_w<18$, the liquid is primarily localised as a droplet in the interior.   

If we now consider increasing the overall system size at partial wetting, whilst maintaining the overall number density constant, then one expects the volume of liquid in the wedge interior to increase in size accordingly. However, the thickness of the exterior lobes should remain approximately unchanged because they arise from the usual accumulation of ABPs at a barrier, combined with a local balance of particles crossing the barrier: these aspects only depend on the liquid density.

\begin{figure}
    \centering
    \includegraphics{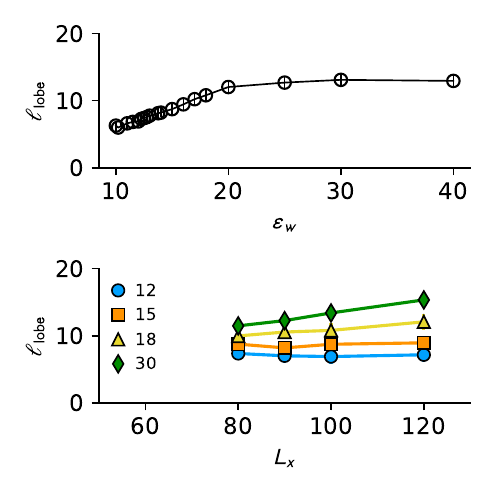}
    \caption{Thickness of the `lobes' ie the adsorbed layer on the exterior of the wedge. The upper panel shows the variation with the barrier strength for a system of size $L_x=100\sigma$. The lower panel shows the scaling with system size for different values of the barrier strength $\epsw$ both in the complete wet regime ($\epsw=30$) and partially wet regime ($\epsw=12$).}
    \label{fig:lobes}
\end{figure}

 Fig.~\ref{fig:lobes}  confirms this: the top panel shows that on increasing the repulsion strength, the thickness of the lobes grows, and eventually saturates as the complete wetting regime is approached.  (This thickness is measured as the stationary average of the largest distance between the outer liquid-vapor interface and the exterior of the wedge).  The lower panel of Fig.~\ref{fig:lobes} shows that in the partially wet regime $\varepsilon<18$, the lobe thickness is independent of system volume. Thus, the exterior lobes in the partial wetting regime can be regarded as a finite-size effect such that for a sufficiently large system, the liquid phase is essentially confined in the wedge interior. 

\subsection{Flow fields}
The internal self-propulsion force renders the system dissipative, promoting flows coupled to the geometric features of the barriers. For example, in the case of ellipsoidal impenetrable barriers, distinctive quadrupolar flow patterns have been reported previously \cite{wagner2022}. 

In order to quantify the flow field, we track the particle displacements $\Delta \mathbf{r}_i=\mathbf{r}_i(t+\Delta t )-{\bf r}_i(t)$ with $\Delta t=0.005\taur$, a time interval that is sufficiently small to allow us to follows the local patterns of motion. We estimate the flow patterns from the $\alpha=x,y,z$ components of the field 
\begin{equation}
    J_j^\alpha (t) = \dfrac{1}{\Delta t \ell^3}\sum_i \Delta{\bf r}_i(t) \cdot{\bf \hat{e}}_{\alpha}\Iell(\bm{r}_i-\bm{b}_j).
\end{equation}
where ${\bf \hat{e}}_{\alpha}$ is the unit vector. We average over the steady state and the $z$ direction to obtain two-dimensional flow maps from which we can extract streamlines. 

Fig.~\ref{fig:polarisation-flux}(e-h) illustrates the flow patterns occurring at weak ($\epsw=12$) intermediate ($\epsw=18$) and strong ($\epsw=30$) and very strong ($\epsw=40$) barriers. In all cases, the wedge geometry shapes the structure of the flow field. Consider first the results for the strongest repulsive barrier ($\epsw=40$). Here the end-points of the wedge correspond to regions of opposing counterflows, reminiscent of the quadrupolar structure of hard objects mentioned above; the flow in the interior of the wedge is from the vapor into the liquid, consistent with the orientation field; particle flow through the wedge occurs only at the wedge corner, where two other counter-flows are formed, and where the flow field rotates in order to be orthogonal to the exterior of the wedge. 

These overall features persist on decreasing $\epsw$, but the flow patterns become more complex. In particular, the flow field reorients itself around the contact points between the droplet and the barrier, forming two additional local regions of nonzero circulation. Softer barriers also engender interesting changes in the flow patterns inside the liquid droplet. One is a marked increase in the flow through the wedge corner. Another is that while for the largest $\epsw$ particles are trapped in the liquid region, with any flow occurring parallel to the barriers (for example, Fig. \ref{fig:polarisation-flux}(h)) in the positive x and y directions, for weaker $\epsw$ the particles can traverse the barrier, creating perpendicular flow patterns as seen in Fig. \ref{fig:polarisation-flux}(e). These changes are accompanied by a reversal in the net flux direction, with $x$ and $y$ flux components being net negative for small $\epsw$ and net positive for large $\epsw$. Forces on asymmetric objects have been previously linked \cite{nikola2016} to the flow patterns of active particles, and the changes observed here indicate a change in the direction of the net force on the wedge as we vary the repulsion strength, a prediction that can potentially be tested in experiments.

These results emphasise that the dissipative non-equilibrium flow patterns generated by the soft wedge can be highly non-trivial, with exquisite features that start to develop even well inside the dilute phase. Notwithstanding these fine-grained phenomena, as we show in the next section, coarse-graining over sufficiently large scales reproduces closely the principal features of the density profiles and their fluctuations that occur in comparable equilibrium systems.

\begin{figure*}[t]
    \centering
\includegraphics[width=\textwidth]{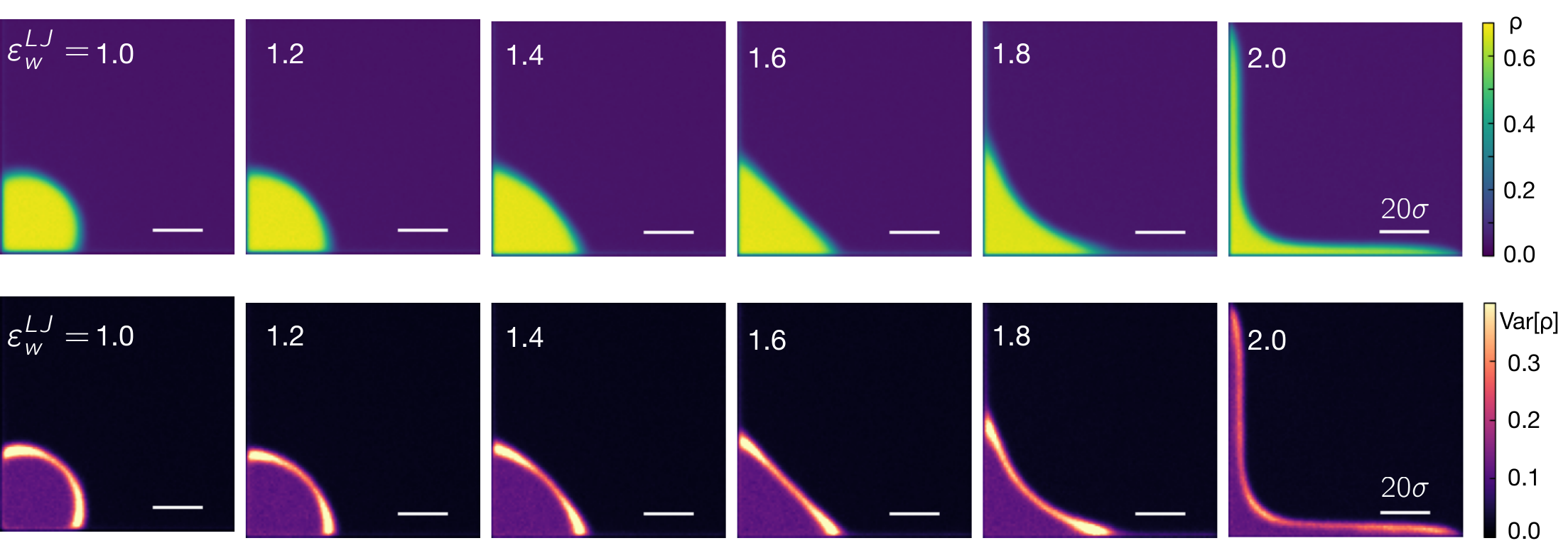}
    \caption{Average and variance of the density profile of the equilibrium LJ system. Upper row: density profile for a droplet of Lennard-Jones liquid in coexistence with the vapour in contact with impenetrable, attractive walls of depth $\epsw$ on the bottom and left sides, and reflective periodic boundary conditions on the top and right sides. Lower row: stationary local density fluctuation profiles for a droplet of Lennard-Jones liquid in coexistence with the vapour in contact with an impenetrable, attractive wall of depth $\epswlj$.}
    \label{fig:lj-densities-var}
\end{figure*}

\begin{figure}
    \centering
    \includegraphics[scale=1]{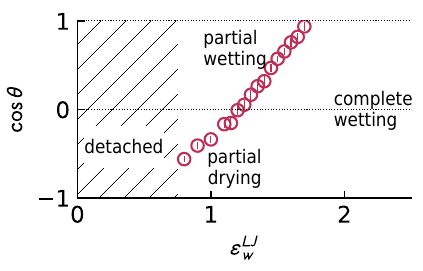}
    \caption{Cosine of the apparent contact angle $\theta$ between the droplet of Lennard-Jones liquid and the 9-3 Lennard-Jones walls  of strength $\epswlj$, as estimated by circular arc fits to the liquid-vapor interface. The hatched area indicates where the droplet detaches from the walls. Note that, differently from Fig.\ \ref{fig:active-cosines}, the range of cosines extends to the partially dry regime ($\cos\theta<0)$ before detaching is observed.
    Error bars are bootstrap estimates for three standard deviations. }
    \label{fig:cosines-lj}
\end{figure}

\section{Comparison with equilibrium wetting in a passive fluid}
\label{sec:lj}

The phenomenology of the active system displays specific spatial density profiles, local density fluctuations and flow patterns. While the non-equilibrium flows have no counterpart in equilibrium, we can compare the stationary density distributions and their fluctuations to a conventional passive system undergoing a wetting transition in an analogous geometry. To do so, we consider the prototypical model for equilibrium wetting in which a liquid droplet of Lennard-Jones particles, in coexistence with the vapor,  interacts with an impenetrable wall supplemented by a long ranged attraction, as described by a Lennard-Jones 9-3 interaction potential.  Such long-ranged wall-fluid attractions are necessary for a planar substrate to engender a first-order wetting transition \cite{evans2019}. We note that for our active system, the competition between attraction and repulsion that engenders a wetting transition arises from a single repulsive potential barrier alone. This contrasts with passive systems, where the necessary ingredients are separate attractive and repulsive parts of the wall-fluid potential.

To make contact with the behaviour of droplets in our active system, we arrange the wall to form a wedge with aperture angle $\alpha=90^\circ$ (see details of the equilibrium model in Appendix~\ref{sec:ljdroplet}). We control the affinity of particles for the wall by varying the strength of the interaction $\epswlj$ between the wedge and the fluid. Note, however, that in contrast to the active case where increasing $\varepsilon_w$ increases the degree of repulsion between the barrier and the fluid particles, here increasing $\epswlj$ strengthens the wall-fluid {\em attraction}. 

Fig.~\ref{fig:lj-densities-var} shows the evolution of the density profile with increasing attractive strength $\epswlj$. The boundary conditions of the simulation box are such that the LJ walls are located on the left and bottom edges while the top and right edges present reflective boundary conditions; the $z$ dimension, orthogonal to the projection plane, is periodic. The qualitative dependence on wall-fluid interaction strength is very similar to that seen for the active case in Fig.~\ref{fig:densities-var}: a droplet is formed in the corner of the wedge, and gradually spreads out as the attractive strength is increased. Complete wetting occurs at the largest wall strengths where the particles coat the wall (note that the rounding of the density profile at the end of the simulation box is a consequence of the reflective boundary conditions required in this ensemble and is not a genuine point of contact). As in the active case, the changes in the density profiles are accompanied by changes in the local density fluctuations, as shown in Fig.~\ref{fig:lj-densities-var}. The regions of highest variance (greatest fluctuation) are located at the vapor-liquid interface, and on approaching the neutral point, $\theta=\pi/2$, the variance exhibits maxima near the point of contact with the wall.

When the attractive wall strength is reduced further,  a situation similar to that observed in the active case arises: the liquid drop pinches off and detaches from the wall.  Fig.~\ref{fig:cosines-lj} plots $\cos(\theta)$ as a function of $\epswlj$ as extracted from circular fits to the interface.  The main difference to the active case is that the passive system can access a larger range of $\cos(\theta)$ before detachment occurs: specifically for the passive system we can stabilize droplets with $\cos\theta\approx-0.5$ ($\theta\approx 2\pi/3$), whereas the active system becomes unstable around the neutral point $\theta=\pi/2$.

\section{Theoretical model for droplet shape fluctuations}
\label{sec:contour}

On sufficiently large (hydrodynamic) length scales, it is natural to describe wetting droplets in a continuum formalism, with a sharp interface separating liquid and vapour domains.   We discuss here a simple model for the fluctuations of such an interface, which is useful for rationalising the density fluctuations observed in active and passive systems.
We describe the interface as a line in two dimensions, in order to model the planar profiles in Fig.~\ref{fig:densities-var}.

\subsection{Contour Model}

\begin{figure}
    \centering
    \includegraphics[width=\columnwidth]{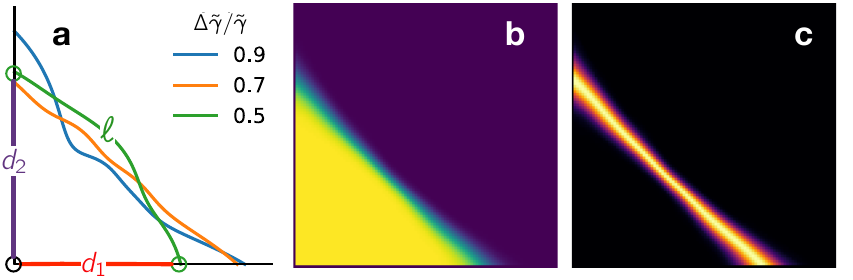}
    \caption{(a) Example paths generated by Monte-Carlo according to the free energy in Eq.~\ref{eq:free-energy}. The quantities $\ell, d_1, d_2$ are illustrated for a specific contour.  (b) Average and (c) variance of the density determined by the statistics of the contours generated at fixed $\Delta \tilde{\gamma}/\tilde{\gamma}=0.7$. For these profiles, no Gaussian convolution is performed. }
    \label{fig:contour-examples}
\end{figure}

For a wedge with opening angle $\alpha$, we describe the interface in polar co-ordinates, with the origin at the corner of the wedge.  The distance from the origin to the interface $R(\phi)$ is parameterised as
\begin{equation}
R(\phi) = R_0 + B\left( \frac{2\phi}{\alpha}-1\right) + \sum_{a=1}^M a_k \sin\left(\frac{\pi k \phi}{\alpha}\right) 
\label{eq:param}
\end{equation}
where $\phi$ is the polar angle with $0\leq\phi\leq\alpha$ and the parameter $M$ enforces a short-wavelength cutoff for numerical convenience.  Example interfaces are shown in Fig.~\ref{fig:contour-examples}a, they are parameterised by $R_0,B,a_1,a_2,\dots,a_M$.  The following results depend weakly on the cutoff $M$.

We formulate our model in terms of the probability distribution for the droplet shape, encoded by the function $R(\phi)$.  Since we describe the interface as a line -- or contour -- we refer to this as the contour model. For simplicity, we suppose that the area of the liquid region is fixed at some reference value $A_0$, so the probability (or probability density) of a given contour is
\begin{equation}
P[R] = \frac{1}{Z} {\rm e}^{-{\cal F}[R]} \delta( A[R]-A_0 )
\label{equ:PR-F}
\end{equation}
where $A[R]=\frac12 \int_0^\alpha R(\phi)^2 d\phi$ is the area enclosed by the liquid-vapour interface, ${\cal F}[R]$ is a free-energy-like quantity that controls the shape fluctuations, and $Z$ is a suitable normalisation constant.

By analogy with equilibrium systems we propose that the dominant terms in ${\cal F}$ on hydrodynamic scales are 
\begin{equation}
{\cal F}[R] = \tilde\gamma_{\rm lv} \ell_{\rm lv}[R] - \Delta\tilde\gamma_{\rm w} d_{\rm wl}[R]
\label{eq:free-energy}
\end{equation}
where $\ell_{\rm lv}$ is the length of the liquid-vapour interface, $d_{\rm wl}$ is the (total) length of the liquid-wall interface, and $\tilde\gamma_{\rm lv},\Delta\tilde\gamma_{\rm w}$ play the role of surface tensions (see below for a more detailed discussion).  One sees immediately that $d_{\rm wl}=R(0)+R(\alpha)$ and the interfacial length is $\ell_{\rm lv}[R]=\int_0^\alpha \sqrt{R(\phi)^2+R'(\phi)^2} d\phi$.

In equilibrium, \eqref{eq:free-energy} is valid on hydrodynamic scales and its parameters are related to the surface tensions between the phases as $\tilde\gamma_{\rm lv}=\gamma_{\rm lv}/(k_{\rm B}T)$ and $\Delta\tilde\gamma_{\rm w}=(\gamma_{\rm wv} - \gamma_{\rm wl})/(k_{\rm B}T)$: here $T$ is the temperature and $k_{\rm B}$ is Boltzmann's constant.\footnote{The notation with tildes is a reminder that the parameters in ${\cal F}$ have the units of inverse length, while surface tensions $\gamma$ (without tilde) have units of energy per length in this $2d$ setup.} In principle other terms might also appear in ${\cal F}$ but these are constrained by locality -- the probability of an interfacial perturbation should not depend on the shape of the interface far away -- and by the symmetries of the problem, and the restriction to large length scales.

For active systems, non-local contributions to ${\cal F}$ cannot be ruled out, but the symmetries of the system (isotropy and translation invariance) suggest that \eqref{eq:free-energy} is a natural starting point for a hydrodynamic theory.  Moreover, for large droplets one sees that $\ell_{\rm lv}$ and $d_{\rm wl}$ are $O(L)$ and $\tilde\gamma_{\rm lv},\Delta\tilde\gamma=O(1)$ which means that shape fluctuations are small, as happens for thermal fluctuations in the thermodynamic limit.  As in that case, the result is that the distribution of shapes is dominated by the most likely contour.  Applying calculus of variations to minimise ${\cal F}$ at fixed droplet area $A_0$ one arrives as Young's equation in the form
\begin{equation}
\tilde\gamma_{\rm lv} \cos\theta = \Delta\tilde\gamma_{\rm w} \; .
\label{equ:tilde-young}
\end{equation}
For completeness, a derivation of this result is given in appendix~\ref{appendix:young-derivation}, which also shows that the curvature of the liquid-vapour interface must be constant everywhere, so that the droplet boundary is an arc of a circle.  Analysing small fluctuations about this most-likely shape also recovers the standard theory of capillary fluctuations, with a spectrum proportional to $q^2/\tilde\gamma_{\rm lv}$.  This suggests that the appropriate $\gamma_{\rm lv}$ that should appear in a Young's equation for active fluids is a capillary surface tension~\cite{fausti2021}, see Sec.~\ref{sec:conc} below for further discussion of this point.

\begin{figure}
    \centering
    \includegraphics[width=\columnwidth]{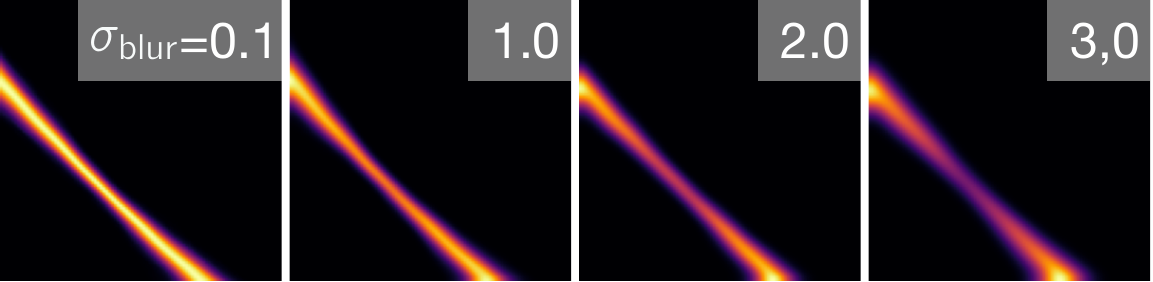}
    \caption{Effect of the Gaussian convolution of scale $\sigma_{\rm blur}=0.1, 1,2,3\delta x$ where $\delta x=0.03 \sqrt{A}$ on the two-dimensional variance profile, at $\Delta \tilde{\gamma}/\tilde{\gamma}=0.7$. }
    \label{fig:var-blur}
\end{figure}

\subsection{Comparison with wetting simulations}

\label{sec:contourcompare}

To investigate the predictions of this model, we sample contour fluctuations numerically by a straightforward Monte Carlo algorithm as detailed in appendix~\ref{app:contour}.
Assuming sharp interfaces, this yields a density field $\rho(x,y|t)$ 
by taking a value of unity for the liquid and zero for the vapor. 
These sharp profiles are then averaged to return the equilibrium density profile, and its variance, 
which are both computed by partitioning the $2d$ domain into bins of size $\delta x$.

\begin{figure*}[t]
    \centering
    \includegraphics{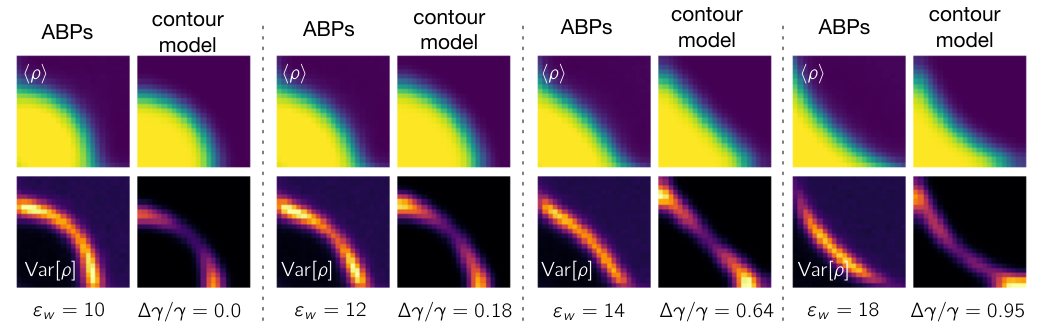}
    \caption{Fitting the density profiles of the active system with the contour model for different values of the barrier strength $\epsw$. The upper row of each panel is the local density, while the lower row is the variance. Every panel displays two columns: the left column corresponds to the data from the region of the active systems contained inside the wedge, excluding the barrier region; the right column represents the fitted profiles produced by an ensemble of 10\ 000 contours. The fit is only performed on the density field, whereas the variance maps are inferred. The numerical values of the barrier strength of the active model $\epsw$ and the fitted values of $\Delta\tilde{\gamma}/\tilde{\gamma}$ are also reported. }
    \label{fig:contourfit}
\end{figure*}

Figure \ref{fig:contour-examples} illustrates the procedure for a particular choice of $\Delta\tilde{\gamma}/\tilde{\gamma}$, connecting collections of paths (Fig.\ref{fig:contour-examples}(a)) to a density profile and its variance (panels (b) and (c)). For a specific choice of $\Delta \tilde{\gamma}/\tilde{\gamma}$ the average density profile converges to form a specific contact angle, such that $\cos\theta = \Delta\tilde{\gamma}/\tilde{\gamma}$: in the specific example of panel (b) and (c) we have $ \Delta\tilde{\gamma}/\tilde{\gamma}=0.7\rightarrow \theta\approx 45^\circ$. While the density profile closely resembles that produced by the passive system (and the interior of the wedge in the active system), the variance profiles are typically sharper and do not display a marked relative increase of the variance in the regions where the contour contacts the wall.

This deficiency of the model arises from our assumption of perfectly sharp instantaneous interfaces. In reality, the density across the instantaneous interfaces (e.g. in the Lennard-Jones system) evolves smoothly, interpolating between the liquid and the vapor density with a characteristic interface width.  (The observed width is also affected by the projection of the three-dimensional system onto a planar density, because of capillary waves along the $z$-direction.)  To account for this width, we perform a convolution of the instantaneous density profiles with a Gaussian kernel of size $\sigma_{\rm blur}$. As illustrated in Fig.~\ref{fig:var-blur}, the convolution promotes the emergence of peaks in local density fluctuations in the vicinity of the contacts with the wedge, similarly to what is observed in both the active and passive particle systems.

It is interesting to quantify the similarities of the fluctuation profiles for the active fluid (Fig.~\ref{fig:densities-var}) and the contour model as a function of the contact angle (or equivalently in the contour model, $\Delta\tilde{\gamma}/\tilde{\gamma}$).  To achieve this, we use Monte Carlo to sample contours with a fixed area $A$. 
First, we calibrate the spatial discretisation of the contour model on the active model in the partially wet regime $\epsw=10$, by tuning the binsize $\delta x$ of the contour model to match the bin location of the contact point in the active system. This gives $\delta x= 0.066 \sqrt{A}$. We fix the discretisation, and by using a blurring scale $\sigma_{\rm blur}=2\delta x$, we compute the average particle density in each bin.  %\emph{** RLJ: what is dx in particle units? **} 
Then, we optimise the parameters of the contour model so that this density profile matches the ABP system at a given state point, by
adjusting $\Delta\tilde{\gamma}/\tilde{\gamma}$ to minimise 
\begin{equation}
{\cal L} = \sum_{{\rm bins}\, j} \left( \bar\rho_j^{\rm ABP} - \bar\rho_j^{\rm contour} \right)^2
\end{equation}
where $\bar\rho_j^{\rm ABP}$ is the average of the local density $\rho_j$ defined as in \eqref{equ:rhoj}, but now scaled to lie between $0$ and $1$; also $\rho_j^{\rm contour}$ is the corresponding (blurred) density for the contour model (which lies between $0$ and $1$ by definition).

Results are shown in Fig.~\ref{fig:contourfit}.
The contour model generates density profiles which closely match those of the ABPs not only for near-neutral conditions (e.g. $\epsw=10$) but also close to the wetting transition $\epsw=18$. The best fit profiles are associated with a contact angle $\cos(\theta)=\Delta\tilde{\gamma}/\tilde{\gamma}$ which confirms the wetting transition to occur around $\epsw=18$ ($\Delta\tilde{\gamma}/\tilde{\gamma}=0.95$) and accords with direct measurements of $\theta$ via circular fits to the interfaces of Fig.~\ref{fig:active-cosines}.   In generating each density profile, we take advantage of the low computational cost of the contour model by performing averages over $10^4$ independent contours.

We also show the local variance of the density.  This is not part of the fitting procedure so it can be interpreted as a prediction of the contour model.  As expected, the variance is large at the liquid-vapour interface.  In the contour model, it also tends to be large where this interface meets the barrier.  For the ABPs under partial wetting conditions, the same behaviour is observed, with large fluctuations near points of contact with the barrier.  However, close to the transition point $\epsw=18$, the strongest density fluctuations in the ABPs occur away from these contact points. In fact, the contour model behaves analogously to the passive Lennard-Jones fluid (recall fig.~\ref{fig:lj-densities-var}), with large fluctuations restricted to the contact points. This suggests that the geometrical setup of the active system may be influencing these density fluctuations, via the tips of the wedge, which are not explicitly accounted for either in the LJ system (which has walls terminating at reflective boundary conditions) or the contour model (which assumes infinite walls). Future work should clarify these potential finite-size effects and their coupling with the nonequilibrium fluxes of Fig.~\ref{fig:polarisation-flux}.

To analyse the density fluctuations in more detail, we characterise the behaviour of the local variance $w(x,y)={\rm Var}[\rho]$, along the interface.  To identify the interfacial region we take two density thresholds $\rho_{\rm lo},\rho_{\rm hi}$ and define a top-hat function $\Pi_{\rho_{\rm lo},\rho_{\rm hi}}(\bar\rho)$ that is equal to unity for $\rho_{\rm lo} < \bar\rho < \rho_{\rm hi}$ (which is the interfacial region) and zero otherwise.  Then we integrate this local variance over the interfacial region to obtain
\begin{equation}
    \overline{w}_{\rm interface}=\int w(x,y) \Pi_{\rho_{\rm lo},\rho_{\rm hi}}[\bar{\rho}(x,y)]dxdy,
\end{equation}
(the integral runs over the entire system and $\bar\rho$ is the local density).

\begin{figure}
    \centering
    \includegraphics[scale=1]{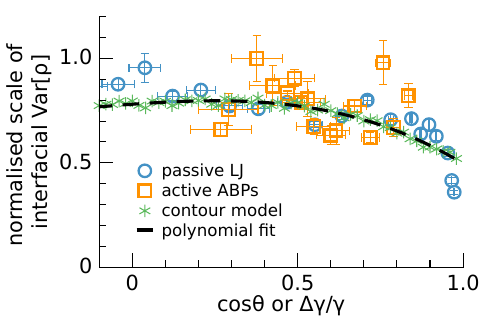}
    \caption{Scaled value of ${\rm Var}[w]_{\rm interface}$ where $w={\rm Var}[\rho(x,y)]$ in the interfacial region between the vapor and the liquid, scaled by its maximum. For the contour model we consider, we convolve the density profiles with convolution scale $\sigma_{\rm blur}=2\delta x$, where $\delta x=0.13 R_0$. 
    }
    \label{fig:varvar-comparison}
\end{figure}

To determine how much $w$ varies as we move along the interface, we compute a corresponding spatial variance:
\begin{equation}
    {\rm Var}[w]_{\rm interface}=\int (w(x,y)-\overline{w}_{\rm interface})^2 \Pi_{\rho_{\rm lo},\rho_{\rm hi}}[\bar{\rho}(x,y)]dxdy.
\end{equation}
We vary the barrier and the wall strength in the ABP model and the (passive) LJ system {respectively}, to obtain systems with various contact angles.
As different models have different scales for the variance $w={\rm Var}[\rho(x,y)]$, we rescale ${\rm Var}[w]_{\rm interface}$ of the active and passive model by their maximal values.  Results are shown in Fig.~\ref{fig:varvar-comparison}, which also shows corresponding quantities for the contour model, computed as a function of $\tilde\Delta\gamma/\tilde\gamma$.   To reduce the noise of our estimates, we repeat the measurements over a number of independent runs, which depend on the complexity of the models (5 for the active system, 20 for the passive system and 100 for the contour model). 

The contour model results can be accurately fitted with a polynomial expression (dashed line), and indicate an increasing spatial variance of the fluctuations, as the system moves from complete wetting [$\cos\theta\approx 1$] towards the neutral point [$\cos\theta\approx 0$]. The particle-based simulations exhibit more pronounced statistical errors, but follow a comparable trend.

Overall, this figure illustrates that -- for a given choice of the contact angle -- and notwithstanding the very different specific interaction mechanisms with the wall/barrier, the three model systems display similar characteristic fluctuations of their interfaces, which match quantitatively (up to an overall scale factor).  =More generally, Figs.~\ref{fig:densities-var}, \ref{fig:lj-densities-var} and \ref{fig:contourfit} demonstrate similar phenomenology for ABPs, passive LJ particles and the contour model, indicating that the principal features of the steady state density distribution in the wedge interior are well described by the minimal contour model, within which the most likely droplet shape is determined by a ratio of surface-tension-like quantities.  This idea is developed further in the next Section.

\section{Discussion and conclusions}
\label{sec:conc}

We have considered active Brownian particles subject to a finite repulsive external potential as a simple model for surface phase transitions in active fluids. Our findings connect the behaviour of stable droplets to previously identified sharp spontaneous symmetry-breaking transitions from completely to partially wet states in three-dimensional systems \cite{turci2021a}. In order to stabilise active droplets with nonzero contact angles, we utilise a wedge-shaped external potential. The droplets thus formed display distinctive density distributions as well as non-trivial steady currents.

A comparison with the wetting behaviour of a (Lennard-Jones) passive fluid confined by impenetrable, long-range attractive walls shows quantitative correspondences between the characteristic features of the density field and its fluctuations. Increasing the strength of the attractive wall-fluid interaction in the passive system plays an analogous role to increasing the potential barrier height in the active system, promoting a continuous change in curvature in the liquid-vapor interfaces and the promotion of density fluctuations in the vicinity of the contact line that increases as one moves from complete wetting ($\cos(\theta)=1$) through the partial wetting regime to the neutral point  ($\cos(\theta)=0$). 

Our results for the partial to complete wetting transition in ABPs provide new insights into phase separation phenomena in dry active matter. They suggest that, despite the inherent non-equilibrium mechanisms that engender phase separation and non-trivial flow patterns, the large-scale properties of the active model can be mapped onto an equilibrium one, an assertion that is supported by the finite-size scaling analysis of the density profiles in the partially wet regime. Such a scenario is similar to the hydrodynamic models proposed to rationalise motility-induced phase separation in the bulk \cite{kourbaneHoussene2018} and connects to recent attempts to recover capillary-wave-like fluctuations in the nonequilibrium case \cite{chacon2022,besse2023,langford2023}. It will be interesting to study whether these findings generalize to more complex active matter systems with anisotropic interactions or long-ranged hydrodynamic effects: in such systems, MIPS may be modified \cite{jayaram2020} or suppressed \cite{theers2018}, hence affecting dramatically one of the key ingredients for surface phase transitions.

The agreement of the contour model with the ABP system supports the correspondence between the interfacial behaviour of MIPS systems and equilibrium fluids.  The parameters $\gama,\gamb$ appearing in $\cal F$ are similar to (rescaled) surface tensions in equilibrium systems in that they set the probabilities of interfacial fluctuations.  
Indeed, one may use (\ref{equ:PR-F},\ref{eq:free-energy}) to write the suggestive expression,
\begin{equation}
    \gama = -\frac{\partial}{\partial \ell_{\rm lv}} \ln P \:,\\
\end{equation}
analogous to the definition of an equilibrium surface tension as a derivative of the free energy.  Similarly $\gamb = (\partial/\partial d_{\rm wl}) \ln P$.
However, it should not be assumed that $\gama,\gamb$ are related to mechanical aspects of interfaces, in contrast to equilibrium systems where the surface tension controls anisotropic contributions to the stress tensor\cite{kirkwood1949}, as well as the magnitude of capillary fluctuations, and the Laplace pressure.  It is a familiar feature of active matter that there is no {\em a priori} connection between fluctuations and mechanical forces, because standard fluctuation-dissipation theorems do not hold in these non-equilibrium steady states.
In deriving Young's equation, the central (albeit trivial) assumption is that the observed large-scale droplet configuration is the one that maximises the probability, so its shape is naturally determined by probabilistic quantities.

To the extent that the contour model is an accurate description of interfacial fluctuations, the quantity $\gama$ can be deduced from fluctuations of a planar MIPS interface: it is closely related to the capillary surface tensions discussed in~\cite{fausti2021,langford2023}.  
In equilibrium, $\gamb$ is the difference between liquid-wall and vapour-wall surface tensions: these can be computed separately from simulations of the fluid in contact with the wall, under appropriate conditions.  In this case, all the quantities that appear in Young's equation are known, and the contact angle $\theta$ can be predicted.  
In the active case, it is not clear how $\gamb$ can be estimated without direct simulation of a wetting droplet, so the modified Young's equation \eqref{equ:tilde-young} cannot \emph{predict} the value of $\theta$.  Instead, one might infer $\gamb$ by measuring the contact angle, under the assumption that (\ref{equ:PR-F},\ref{eq:free-energy}) are suitable as a model for droplet-shape fluctuations.  Under this assumption, the theory does make non-trivial predictions, for example that the contact angle of a fluid in a wedge should be independent of the aperture angle $\alpha$, as found (at least approximately) in Fig.~\ref{fig:apertures}.  Predictions of the most likely shape and its fluctuations would then also available for other wall geometries.  However, the existence of steady-state currents in active fluids should be borne in mind since this could mediate non-local interaction terms in ${\cal F}$, in which case (\ref{equ:PR-F},\ref{eq:free-energy}) would not hold.  More detailed tests of this theory would be desirable.

 Our measurements of $\cos(\theta)$ as a function of the wall-fluid repulsive interaction strength (fig.~\ref{fig:active-cosines}) show the wetting transition in ABPs to be first order. This finding accords with that of our previous work \cite{turci2021a} in which the transition was studied using very different methods. Interestingly, however,  in equilibrium fluids first order wetting is associated with the presence of long-range wall-fluid interactions \cite{evans2019}, while in our ABP simulations the repulsive barrier potential Eq.~(\ref{eq:barrierpot}) is intrinsically very short ranged. It is possible that {\em effective} long-ranged interactions may be generated in active matter via many-body correlations and long-ranged flow fields; this is believed to occur at least for the case of a disordered boundary~\cite{bendor2022}. Investigating the matter of what controls the order of the wetting transition in active wetting would be an interesting topic of further work. Other directions suggested by our research include 
 a better understanding of the differences between three and two-dimensional active wetting, and the quantification of finite-size effects on the magnitude of the contact angles.

\appendix
\section{Active Brownian Particles model}
\label{sec:model-details}
We follow previous literature and simulate repulsive active Brownian particles of equal mass $m$ interacting via the Weeks-Chandler-Anderson potential

\begin{equation}
  V(r)=4 \varepsilon\left[\left(\frac{\sigma}{r}\right)^{12}-\left(\frac{\sigma}{r}\right)^{6}\right]+\varepsilon
\end{equation}
with a cutoff at $r= 2^{1/6}\sigma$.

The equations of motion for the particle positions and orientations  are: 
\begin{align}
\partial_{\mathrm{t}} \mathbf{r}_{\mathrm{i}} &=\beta D_{\mathrm{T}}\left[\mathbf{F}_{\mathrm{i}}+F_{\mathrm{p}} \mathbf{p}_{\mathrm{i}}\right]+\sqrt{2 D_{\mathrm{T}}} {\Lambda}_{\mathrm{r}}\:,  \label{eq:rmotion} \\
\partial_{\mathrm{t}} \mathbf{p}_{\mathrm{i}}&=\sqrt{2 D_{\mathrm{R}}}\left(\mathbf{p}_{\mathrm{i}} \times {\Lambda}_{\mathrm{p}}\right).\label{eq:pmotion}
\end{align}
where $ \mathbf{p}_i$ is the orientation of particle $i$, $F_p$ is the strength of self propulsion force $D_{\mathrm{T}}$ and $D_{\mathrm{R}}$ are diffusivities and $\Lambda_{\rm r}$ and $\Lambda_{\rm p}$ are noise terms.

The translational and rotational diffusion constants $D_{\mathrm{T}}$ and $D_{\mathrm{R}}$ coupling is $D_{\mathrm{T}}=D_{\mathrm{R}}\sigma^2/3$, with inverse thermal energy scale $\beta$ and friction $\xi$ such that $\beta\xi=1/D_{\rm t}$. Following Stenhammar et al.\cite{stenhammar2014}, we keep the self-propulsion force constant $F_p = 24\varepsilon/\sigma$ as well as the friction $\xi=50 \sqrt{\varepsilon m}/\sigma$. The noise terms ${ \Lambda}_{\mathrm{r}},{\Lambda}_{\mathrm{\theta}}, {\Lambda}_{\mathrm{p}}$ are unit-variance stochastic vectors in three dimensions whose Cartesian components satisfy $\left\langle\Lambda_{\mathrm{i}}(\mathbf{r}, t) \Lambda_{\mathrm{j}}\left(\mathbf{r}^{\prime}, t^{\prime}\right)\right\rangle=\delta_{\mathrm{ij}} \delta\left(\mathbf{r}-\mathbf{r}^{\prime}\right) \delta\left(t-t^{\prime}\right)$. The rotational diffusion constant defines a natural timescale for the system, the rotational diffusion time $\tau_R = 1/D_R$. We work at constant P\'eclet number $\Pe=F_{\rm p}/\xi D_{\rm r}\sigma=60$. We choose this value as a compromise between the need to work away from criticality ($\Pe_{\rm}\approx 36$ \cite{turci2021}) and the requirement of system sizes that are comparable or larger than the persistence length $\ell_{\rm p}=\sigma\Pe$.

To integrate the equations of motion, we implement an Euler-Maruyama scheme with constant timestep $dt=4\cdot 10^{-5}\tau_R$,  following the Ermak-McCammon method described in detail in \cite{das2018}, with an in-house implementation for the molecular dynamics package LAMMPS \cite{plimpton1995}.\\

\section{Lennard-Jones droplet}
\label{sec:ljdroplet}
We performed molecular dynamics simulations of a system of point-like particles interacting via the short-range Lennard-Jones potential

\begin{equation}
V(r) = 4 \varepsilon_{LJ}\left[\left(\frac{\sigma}{r}\right)^{12}-\left(\frac{\sigma}{r}\right)^{6}\right]\:.
\end{equation}
The potential was truncated and shifted at cutoff $r_c=2.5\sigma$ .

To mimic the setup of the active case, we considered a three-dimensional system with $L_z\ll L_x=L_y$, with $L_z=20\sigma$ and $L_x=L_y = 100\sigma$. While the $z$ dimension is periodic, particles interact with a Lennard-Jones 9-3 wall interaction on two of the remaining faces (forming a corner), while the opposite walls have reflective boundary conditions. The wall-fluid interactions took the form
\begin{equation}
V_w(r) = \epsilon_w^{LJ} \left[\frac{2}{15}\left(\frac{\sigma}{u}\right)^9-\left(\frac{\sigma}{u}\right)^3\right],
\end{equation}
where $u$ is the distance from the wall. Wall-fluid interactions were cutoff at $r_{\rm cut}^{w}=L_x-0.5\sigma$.

The simulations were performed in the NVT ensemble at temperature $T=0.919 54\varepsilon_{LJ}$  for a system of $N=20\ 000$ particles of mass $m$ and  number density $\rho=0.1\sigma^{-3}$ using a timestep $dt=0.006\sqrt{m\sigma^2/\varepsilon_{LJ}}$. 

\section{Fitting contact angles}

\label{sec:contactangle}
Measuring contact angles relies on fitting procedures. To validate our results, we follow two alternative procedures to fit the contact angles from stationary density profiles of the active and passive models: (1) a three-parameter fit with a circular profile; (2) a seven-parameter fit with a Fourier series representing the liquid-vapor interface.

For method (1), we identify the interface between vapor and liquid by thresholding the absolute value of the gradient of the density profile to extract the ($x$,$y$) coordinates of points exclusively at the vapor-liquid. We ignore interfacial points close to the walls as they worsen the quality of the result. We then perform a least-square circular fit with parameters $R,x_c,y_x$ where $R$ is the radius of the circle  $x_c$ and $y_c$ are the coordinates of the centre of the circle. The centre of the circle corresponds to the tip of the wedge only when the contact angle is $90^\circ$. In general, the wedge identifies the sector of a circle with the centre inside or outside of the wedge. When the centre is outside, the contact angle is measured as $\theta = \pi/2-\arcsin{(|y_c|/R)}$, when it is inside, the angle is $\theta = \pi/2+\arcsin{(|y_c|/R)}$.

For method (2), we use the parametrisation of Eq.~\ref{eq:param} with $M=5$ to fit the interface between the vapor and the liquid. In this case, we include all the points of the interface that are inside the wedge and evaluate the local slope of the interface at a fixed distance $d_\theta = 3\sigma $ from the wall in order to approximate the contact angle. We take advantage of the closed form parameterisation (Eq.~\ref{eq:param}) by taking its derivative analytically, which improves the numerical stability of the estimate of the contact angle.

\begin{figure}
    \centering
    \includegraphics{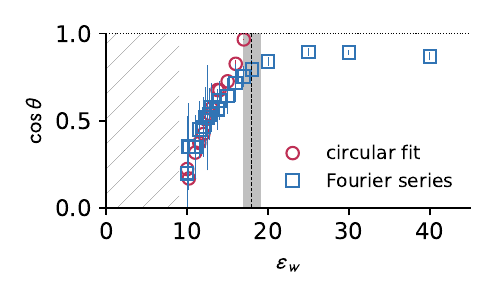}
    \caption{Comparison of the two independent measures of contact angles in the case of the active droplets. The hatched area indicates the detached regime.}
    \label{fig:comparison-cosines}
\end{figure}

The two methods yield comparable results in the partially wet regime. Method 2 breaks down on the approach to the wetting transition, although it continues to provide a lower bound for the contact angle at high $\epsw$. This is illustrated in Fig.~\ref{fig:comparison-cosines}. 

\section{Variational derivation of Young's equation}
\label{appendix:young-derivation}

\begin{figure}
    \centering
    \includegraphics[scale=1]{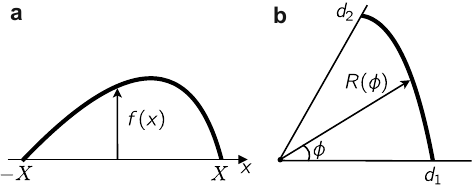}
    \caption{(a) Sketch of geometry for derivation of Young's equation in Appendix~\ref{appendix:young-derivation}.  (b) Corresponding geometry for the contour model, see Appendix~\ref{app:contour}.}
    \label{fig:sketch-coord}
\end{figure}

We show that Young's equation can be obtained by minimising the free-energy-like quantity \eqref{eq:free-energy}, subject to a constraint of fixed enclosed area. We emphasize however than in the active case ${\cal F}$ is not a thermodynamic free energy, instead it describes the log-probability of shape fluctuations, such that its minimum is the most likely shape.  For convenience we consider a planar wall instead of a wedge geometry and define $f(x)$ as the normal distance from the wall to the interface, see  Fig.~\ref{fig:sketch-coord}.  We explain below how results for other geometries can be deduced from this computation.  (In order for $f$ to be single-valued our derivation also requires that $\theta\leq\pi/2$, this assumption can be also be relaxed.)  We assume that the interface meets the wall at $x=\pm X$ so $f(X)=0=f(-X)$: here $X$ is a variational parameter.   The length of the liquid-vapour interface is $\ell_{\rm lv} = \int_{-X}^X \sqrt{1+f'^2} dx$ and the length of the liquid-wall interface is $2X$.   The enclosed area is $\int_{-X}^X f dx$.  

The problem can be simplified by a symmetry argument: the optimal droplet shape is symmetric about $x=0$ so $f'(0)$.
We enforce the constraints of enclosed area $A_0$ and $f(X)=0$ by Lagrange multipliers $\lambda,\mu$.  Hence it remains to extremise
\begin{equation}
{\cal L}[f,X] = 2 \int_{0}^X \left[ \gama \sqrt{1+f'^2} - \lambda f \right] dx - 2 \gamb X + \lambda A_0 - \mu f(X) 
\end{equation}
(Note that this is a functional of $f$, also $X$ is a scalar variational parameter and $A_0$ is the imposed area.)

Within the calculus of variations we write $f=f+\delta f$ and $X=X+\delta X$.  Substituting this into ${\cal L}$, the optimal droplet shape is identified by setting the first variation $\delta {\cal L}=0$.  We find 
\begin{align*}
\delta {\cal L}&={2}\int_{0}^{X}\left[\gama \frac{f^{\prime}}{\sqrt{1+f^{\prime 2}}} \delta f^{\prime}-\lambda \delta f\right] d x 
\\& \qquad +\delta X\left[ {2} \gama \sqrt{1+f^{\prime}(X)^{2}}- {2}\lambda f(X)-{2}\gamb-\mu f^{\prime}(X)\right] \\ 
& \qquad -\mu \delta f(X)
\end{align*}
Integration by parts and using $f(X)=0$ and $f^\prime(0)=0$ yields 
\begin{align}
\delta {\cal L}&=\left[{2}\gama \frac{f^{\prime}(X)}{\sqrt{1+f^{\prime}(X)^{2}}}-\mu\right] \delta f(X)
\nonumber\\&\qquad -{2}\int_{0}^{X}\left[\gama \frac{d}{d x}\left(\frac{f^{\prime}}{\sqrt{1+f^{\prime 2}}}\right)+\lambda\right] \delta f d x
\nonumber \\&\qquad +\delta X\left[ {2}\gama \sqrt{1+f^{\prime}(X)^{2}}
-{2}\gamb-\mu f^{\prime}(X)\right]
\label{equ:to-min}
\end{align}
The optimal droplet has $\delta L=0$ for any $\delta f,\delta X$, so all the terms in square brackets need to vanish. 
The second of these terms yields an Euler-Lagrange equation for $f$ whose solutions are circular arcs with radius $1/|\gama\lambda|$.\footnote{In equilibrium, $\lambda$ is related to the Laplace pressure.}  To obtain the contact angle one must deal with the end points of these arcs, for which
the first term in \eqref{equ:to-min} implies that
\begin{equation}
\mu={2}\gama \frac{f^{\prime}(X)}{\sqrt{1+f^{\prime}(X)^{2}}}
\end{equation}
Putting this into the third term, we get (after simplification)
\begin{align}
\gamb
%&=\gama \left[\sqrt{1+f^{\prime}(X)^{2}}-\frac{f^{\prime}(X)^{2}}{\sqrt{1+f^{\prime}(X)^{2}}}\right]\\
 &=\gama \frac{1}{\sqrt{1+f^{\prime}(X)^{2}}}
 %\\&=\gamma_{1} \cos \theta
\end{align}
Finally note that $f'(X)$ is the gradient with which the liquid-vapour meets the wall, which is $\tan\theta$, where $\theta$ is the contact angle.  We assumed $0<\theta\leq\pi/2$ so $1/\sqrt{1+f^{\prime}(X)^{2}}=\cos\theta$ and we recover Young's equation in the form \eqref{equ:tilde-young}.
%which is Young's equation.

Note that we performed this computation for a particular parameterisation of a droplet on a planar substrate, but the condition that the terms in square brackets vanish in \eqref{equ:to-min} gives \emph{local} geometrical constraints on the shape of the optimal droplet, which can be formulated in terms of the local curvature and the contact angle, independent of the parameterisations. As a result, the geometrical properties of the minimiser can be transferred to other geometrical settings even if the natural parameterisations are different in that case [recall~\eqref{eq:param}].  Specifically, the optimal shape has sections of liquid-vapour interfaces that form arcs of circles, and Young's equation is obeyed at points of contact with (locally) planar substrates.

\section{Contour sampling}
\label{app:contour}

We describe a numerical procedure for sampling droplet shapes according to \eqref{equ:PR-F}, subject to a constraint of fixed area $A[R]=\frac12 \int_0^\alpha  R(\phi)^2 d\phi$.  To achieve this note that
\begin{equation}
A[R] = \frac{\alpha}{2} \left[ R_0^2 + R_0 \sum_{k {\rm \,odd}} \frac{4 a_k }{k\pi} +  \frac{B^2}{3} + \frac{1}{2}\sum_k a_k^2  - B \sum_{k {\rm \,even}} \frac{4a_k }{k\pi} \right]
\end{equation}
Given parameters $(B,a_1,\dots,a_M)$ and a target area $A_0$ this is a quadratic equation that can be solved for $R_0$.  We fix $R_0$ in this way and perform Metropolis MC on $\mathbf{u}=(B,a_1,\dots,a_M)$.\footnote{Note that the interpretation of \ref{equ:PR-F} as a probability density for functions $R$ has some ambiguity, but the most likely droplet shape is unambiguous and obeys Young's equation.  Our choice of MC sampling method corresponds to a specific interpretation of the probability density $P[R]$.}
As MC updates we propose 
\begin{equation}
    \mathbf{u}_{\rm new} = \mathbf{u}_{\rm old}+\mathcal{N}
\end{equation}
where $\mathcal{N}$ is a vector of Gaussian random numbers where each component has mean zero and variance $s^2$.  We take $s=0.01$.
The proposed update is accepted with probability ${\rm min}(1,{\rm e}^{{\cal F}_{\rm old}-{\cal F}_{\rm new}})$, evaluation of ${\cal F}$ requires the integral $\ell_{\rm lv}[R]=\int \sqrt{R(\phi)^2+R'(\phi)^2} d\phi$, which is performed numerically.  

Note that scaling the droplet as $R(\phi)\to\lambda R(\phi)$ for any scale factor $\lambda$, we find that $A\to\lambda^2 A$ and ${\cal F}\to\lambda{\cal F}$.  This scale invariance means that there are only two non-trivial parameters in this sampling problem, which are $\gamb/\gama$ and $\gama^2 A_0$.  For numerical purposes we therefore fix $A_0=1$ without loss of generality.

For the range of $\Delta\tilde{\gamma}/\tilde{\gamma}$ here considered we observe (after an initial transient) decorrelation of ${\cal F}$ over approximately MC 50 steps, which allows us to accumulate large statistics. For example, we produce $200\ 000$ independent profiles to evaluate the density and variance of the density field in blocks of 2000 contours, to produce 100 samples of the variance in the interface region, from which we estimate the scale of variance fluctuations within the interface (Fig.~\ref{fig:varvar-comparison}).

\section*{Author Contributions}

 All authors were involved in conceiving and directing the research. FT performed the numerical calculations. All authors wrote the paper. 

\section*{Conflicts of interest}
There are no conflicts to declare.

\section*{Acknowledgements}
We thank R.~Evans for a helpful conversation on the role of fluctuations in surface phase transitions and M. Cates for discussions of interfacial properties of active fluids. The computer simulations were carried out using the computational facilities of the Advanced Computing Research Centre, University of Bristol, as well as the Isambard 2 UK National Tier-2 HPC Service (\href{http://gw4.ac.uk/isambard/}{http://gw4.ac.uk/isambard/}) operated by GW4 and the UK Met Office, and funded by EPSRC (EP/T022078/1).

%%%END OF MAIN TEXT%%%

%The \balance command can be used to balance the columns on the final page if desired. It should be placed anywhere within the first column of the last page.

%If notes are included in your references you can change the title from 'References' to 'Notes and references' using the following command:
%\renewcommand\refname{Notes and references}

%%%REFERENCES%%%
% \bibliographystyle{apsrmp}
%\bibliography{biburl,papers,extra}

%apsrev4-2.bst 2019-01-14 (MD) hand-edited version of apsrev4-1.bst
%Control: key (0)
%Control: author (8) initials jnrlst
%Control: editor formatted (1) identically to author
%Control: production of article title (0) allowed
%Control: page (0) single
%Control: year (1) truncated
%Control: production of eprint (0) enabled
%

\end{document}